\documentclass[aps,prd,preprint,groupedaddress,showpacs]{revtex4}
\usepackage{epsf,epsfig,graphics,graphicx}
\usepackage{verbatim,color,ulem}
\newcommand{\be}{\begin{equation}}
\newcommand{\ee}{\end{equation}}
\newcommand{\bea}{\begin{eqnarray}}
\newcommand{\eea}{\end{eqnarray}}
\newcommand{\ba}{\begin{array}}
\newcommand{\ea}{\end{array}}

\begin{document}
\title{Holographic Approach to Nonequilibrium Dynamics of Moving Mirrors Coupled to Quantum Critical Theories}
\author{Chen-Pin Yeh}
\email{chenpinyeh@mail.ndhu.edu.tw}
\author{Jen-Tsung Hsiang}
\email{cosmology@gmail.com}
\author{Da-Shin Lee}
\email{dslee@mail.ndhu.edu.tw}

\affiliation{Department of Physics, National Dong-Hwa University,
Hualien 97401, Taiwan, Republic of China}

\begin{abstract}
We employ the holographic method to study  fluctuations and
dissipation of an $n$-dimensional moving mirror coupled to quantum
critical theories in $d$ spacetime dimensions. The bulk
counterpart of the mirror with perfect reflection is a
$n+1$-dimensional membrane in the Lifshitz geometry of $d+1$
dimensions. The motion of the mirror can be realized from the
dynamics of the brane at the boundary of the bulk. The excited
modes of the brane in the bulk render the mirror undergoing
Brownian motion. For small displacement of the mirror, we derive
the analytical results of the correlation functions and response
functions. The  dynamics of the mirror due to small fluctuations
around the brane vacuum state in the bulk is found supraohmic so
that after
 initial growth, the velocity fluctuations approach a saturated
value at late time with a power-law behavior. On the contrary, in
the Lifshitz black hole background, the mirror in thermal
fluctuations shows that its relaxation dynamics becomes ohmic, and
the saturation of velocity fluctuations is reached exponentially in
time. Finally a comparison is made with the result of a moving mirror driven by free fields.
\end{abstract}


\maketitle
\section{Introduction}
Understanding of the microscopic origin of dissipation and
fluctuations in a nonequilibrium system is the main concern in
statistical mechanics. One of the ubiquitous nonequilibrium
phenomena in nature is Brownian motion in which an object moves
under a fluctuating environment. In this case, the Langevin
equation is known to provide a satisfactory description of the
Brownian particle, and its generic form  is given by
 \be
  m \ddot{X}(t)+ \int dt' \,  \theta(t-t') \,
  \mu (t-t')X(t')=R(t) \, , \label{langevin_eq}
  \ee
where $X(t)$ is the position of the particle. This Langevin equation
is a classical equation of motion modified phenomenologically by two
terms that incorporate both dissipation and fluctuation effects upon
the particle in a random medium. The
memory kernel $\mu (t)$ accounts for the dissipation effect, and
in general depends on the past histories of the particle. The noise force
$R(t)$ that mimics the random environment and is correlated
over time scales determined by the typical scales of the medium. The
statistic properties of the noise are specified by
 \be
 \langle
R(t)\rangle=0 \,, \qquad\qquad\langle R(t)R(t')\rangle=\eta(t-t')
\, . \ee These two effects are ultimately crucial for the system to
evolve into thermodynamic equilibrium of the Brownian particle with
the environment, and are thus related by the fluctuation-dissipation
theorem.

A microscopic description that leads to the Langevin equation~(\ref{langevin_eq})
has been developed by Caldeira and Leggett~\cite{Leggett} within the
context of one-particle quantum mechanics. The idea is to consider a
specific system-environment model where the particle interacts
bilinearly with an environment.  The effects of environmental
degrees of freedom on the particle can be summarized with the
method of Feynman-Vernon influence
functional~\cite{Fv} by integrating out environment variables. In
the classical approximation where the intrinsic quantum
uncertainties of the particle is ignored, the Langevin equation can
be obtained by minimizing the corresponding effective stochastic
action. This effective action can be exactly derived if the
environment variables are Gaussian and their
coupling with the system is linear~\cite{GSI,Hu_92}. However many
interesting strongly coupled environments, such as
condensed matter systems, are out of reach by the method of
influence functional.

The study of the Brownian
particle has been extended to the nonequilibrium
dynamics of a charged oscillator in nontrivial quantized
electromagnetic-field backgrounds~\cite{Lee_08,Lee_12} and to a
perfectly reflecting mirror moving in a quantum field~\cite{Lee_05,Lee_13}. The moving mirror
problem is of interest in its own right. For example, this problem
can be related to the dynamical Casimir effect. When the mirror
undergoes nonuniform acceleration, it is expected to create quantum
radiation that in turn damps out the motion of the mirror as a
result of the motion-induced radiation reaction force. In 3+1-dimensional spacetime, this problem is solved
 only for small mirror displacement. The force acting on
the mirror is the radiation pressure of the environmental field
that arises from the area integral of the stress
tensor. Then the coarse-grained effective action is obtained by
integrating out the quantum field, and thus the corresponding
semiclassical Langevin equation of the form (\ref{langevin_eq}) is
derived. In the small displacement and slow motion limit of the
mirror, the emission of quantum radiations that accounts for damping
is found hardly detectable in a quantum vacuum environment. It is
still unclear whether or not this effect can be observed in a
strong coupling environment.

To understand these issues, in this paper we plan to employ
holographic duality to study the nonequilibrium dynamics of a
mirror moving in a quantum/thermal bath. The prototype of the
holographic duality is the AdS/CFT correspondence, which is a
weak-strong coupling duality between the type IIB string theory in
AdS background and the $\mathcal{N}=4$ super Yang-Mills
theory~\cite{AdSCFT}. It is soon generalized to other backgrounds
and field theories, and has been proven fruitful in applying to
the strong coupling problems in condensed matter systems and the
hydrodynamics of the quark-gluon plasma etc. The first
investigations in applying AdS/CFT to study the dissipation
behavior in the strongly coupled field theory were done
independently in \cite{Herzog:2006gh,Gubser_06,Teaney_06}, where
the probed particle is represented by a string hanging from the
boundary of the AdS black hole. Later progresses have been made to
understand the fluctuations of this end point of the string as
Brownian
motion~\cite{Son:2009vu,Giecold:2009cg,Myers:2007we,CasalderreySolana:2009rm,Huot_2011}.
For a general review on application in the nonequilibrium
dynamics, see \cite{Holographic QBM}. In recent
studies~\cite{Tong_12}, Tong and Wong have discussed  Brownian
motion in the environments at their quantum critical points using
the holographic duality. The quantum critical point is a fixed
point theory with the following scaling symmetry:
  \be
  t\rightarrow\Lambda^zt \, ,\qquad\qquad x\rightarrow\Lambda x \, .
  \ee
For $z=1$, it is the usual scale invariance from conformal
symmetry. Other values of $z$ can arise from finite temperature multicritical points ($z=2$ for
the Lifshitz point) of the condensed matter systems.  Quantum critical
points can also be realized in the strongly correlated electron system,
for example, the dimer model~\cite{dimer model}. The holographic dual
for such quantum critical theories has been proposed
in~\cite{Kachru_08}, where the gravity theory is in the Lifshitz
background:
 \be
 \label{Lifshitz geometry}
 ds^2=L^2\bigg(-r^{2z}dt^2+r^2d\vec{x}^2+\frac{dr^2}{r^2}\bigg) \, .
 \ee
In the following, the curvature radius $L$ is set to be unity.
In~\cite{Tong_12}, the authors derived the analytic results of
dissipation and random force correlation functions in quantum
critical theories via holographic duality in the Lifshitz
background.

The idea of this paper is to model the problem of a mirror moving
under quantum critical theories in terms of the gravity theory with
scaling symmetry characterized by the exponent $z$ via the
holographic duality. We can then compare the results with the findings
in \cite{Lee_05} for $z=1$ in the case of the relativistic quantum
field, and further generalize it to other quantum field theories for
$z\neq 1$. In the next section, our
holographic setup will be explained in details, and the response
function to an external force on the two-dimensional mirror will be derived. In
Sec. III, we generalize to a mirror
of general dimensions and also compute the response function and the
correlation function that exhibit essential properties of Brownian
motion. The fluctuation-dissipation theorem is then verified. In
Sec, IV, the finite temperature
Brownian motion of a moving mirror in the holographic setup will be
studied. We then conclude and point out some future works in Sec. V.

\section{Holographic setup for the moving mirror problem}
In this section we propose a holographic setup for coupling a
mirror to quantum critical theories. Here we employ the
``bottom-up" method for the holography duality, where we leave the
derivation of the duality in general backgrounds as an open
question but are content in assuming that there is a field theory
dual to the gravity setup we consider here. The gravity
counterpart for the two-dimensional mirror is a three-dimensional
membrane (3-brane). We consider it as a probed brane moving in the
$d+1$-dimensional Lifshitz background with the metric given in
($\ref{Lifshitz geometry}$). Here the nature of this 3-brane is
left unspecified, while the field theory dual to Lifshitz geometry
is still unclear (see \cite{Kachru_08} and the follow-up papers).
But we expect the behaviors we found in this paper can be general
for large classes of strongly coupled field theories. Let us
consider another $d-1$-brane extended in all spatial directions
other than $r$ and located at $r=r_b$. We would like to interpret
this as a boundary brane where the boundary theory lives. And
$1/{r_b}$ can be regarded as the UV cutoff in the boundary
theory~\cite{Tong_12}, and its physical interpretation will be
given later.
 Let the probed 3-brane end on the boundary $d-1$-brane and extend in the
$x_1$ and $x_2$ directions. We then treat these two directions as
spatial directions of a two-dimensional mirror and other spatial
directions $x^I$, with $I=3,4,\ldots,d$ as perpendicular directions
to the mirror's surface. In the bulk, the position of the 3-brane
can be parametrized (in the static gauge) by $x^I(t,r,x_1,x_2)$. We
assume a rigid mirror, so $x^I$ is independent of $x_1$ and $x_2$.
The 3-brane is governed by the DBI action
$S_{DBI}=-T_3\displaystyle\int dr \, dt \, dx_1 \, dx_2 \,
\sqrt{\det h_{ab}}$, where $h_{ab}$ is the induced metric on the
brane and $T_3$ is the brane tension. Here, for simplicity, we have
assumed the trivial dilaton background and turn off the gauge fields
on the 3-brane. In the Lifshitz geometry it reduces to the following
action:
  \bea
  S_{DBI}&=&-T_3\int dr \, dt \, dx_1 \, dx_2 \,
  r^{z+1}\sqrt{1+r^4x'^I x'^I-\frac{\dot{x}^I\dot{x}^I}{r^{2z-2}}+\frac{(x'^I\dot{x}^I)^2}{r^{2z-6}}- \frac{( x'^I x'^I)(
\dot{x}^J\dot{x}^J)}{r^{2z-6}}} \,
   \nonumber\\
  &\approx& {\rm constant}- \frac{T_3}{2}\int dr \, dt \, dx_1 \, dx_2 \,
\bigg( r^{z+5} x'^I x'^I- \frac{\dot{x}^I\dot{x}^I}{{r^{z-3}}}\bigg)
\, ,
  \label{s_dbi}
  \eea
where $x'^I =\partial_{r} x^I$ and $\dot{x}^I=\partial_{t} x^I$ and
the last expression in (\ref{s_dbi}) is obtained by assuming small
variations of $x^I$ around the minimal energy configuration. Since
all modes in the $I$ directions are independent, we assume that the
motion of the mirror along one of them denoted by $x$. The equation
of motion for the expectation value
of $x$
 in frequency space becomes
 \be
 \label{mirror eom}
 \frac{\partial}{\partial r}\bigg(r^{z+5}\frac{\partial  }{\partial
 r} \langle x\rangle\bigg)+\frac{\omega^2}{r^{z-3}}\langle x \rangle=0 \, .
 \ee
In \cite{Lee_05,Lee_13}, we investigate the dynamics of a perfectly
reflecting mirror when it couples with the quantum field. Their
mutual coupling can be derived from the Dirichlet boundary conditions of the field we imposed
on the mirror. In the field-theoretic approach, the effective coupling in the limit of small
displacement is shown to take the form
  \be
  \label{coupling}
\int dt \, F (t) \, X (t) \, ,
  \ee
where $X$ is the position of the mirror. The radiation pressure of the field $F(t)$ on the mirror is given by the
expectation value of the energy momentum tensor in either the vacuum or the thermal state of the field
  \be
  F (t)=\int dx_1 \, dx_2 \,\langle T_{x, x} \rangle \, .
  \ee
Here $T_{x, x}$ is the component of the energy momentum tensor of
in the direction of
 mirror's motion. It is found that the force arising from the quantum field
 cannot be evaluated infinitesimally
 close to the surface of
the mirror  due to short-distance divergences~\cite{Lee_05}, which
later can be resolved by introducing a fluctuating
boundary~\cite{FS}. Thus, the introduced $1/r_b$, a short-distance
scale, naturally gives uncertainties of the location of  the
mirror's surface at $r=r_b$.

In the holographic setup, the variable \be  X (t)=x (t,r=r_b)
\label{X_x} \ee is the boundary value of the 3-brane position.
Here we assume the boundary of the probed 3-brane can have an
effective coupling like the one in (\ref{coupling}). Additionally,
we would like to emphasize that although this type of the coupling
is obtained for the problem of a moving mirror influenced from
radiation fields, our following holographic approach is also
applied for an extended object as long as its coupling to quantum
critical theory can be described by (\ref{coupling}).
 Varying the action gives
  \be
  \label{mirror boundary}
  T_3S\,r_b^{z+5}\frac{\partial \langle x \rangle }{\partial r}\bigg|_{r=r_b}=F
  \ee
where $\langle x \rangle$ satisfies the equation of motion
(\ref{mirror eom}) and $S$ is the area of the mirror. Similar to the
study of the Brownian particle in \cite{Tong_12}, we calculate the
response function  in this holographic setup by first solving
(\ref{mirror eom}) with the incoming-wave boundary condition, which
is a usual holographic prescription for the retarded Green
function~\cite{Tong_12}. Then the solution to the equation of motion
is the Hankel function of the first kind
  \be
  \langle x (t,r) \rangle  =\frac{1}{r^{2+\frac{z}2}}H^{(1)}_{\frac2z+\frac12}\bigg(\frac{\omega}{zr^z}\bigg)e^{-i\omega
  t} \, .
  \ee
The force acting on the mirror can be calculated  by (\ref{mirror
boundary}):
   \be
   F(\omega)=T_3S\omega r_b^{-\frac{z}2+2}H^{(1)}_{\frac2z-\frac12}\bigg(\frac{\omega}{zr_{b}^z}\bigg)e^{-i\omega
   t} \, .
   \ee
The linear response to  this force is described by
  \be
 \langle X(\omega)\rangle=\chi(\omega, z)F(\omega) \, .
  \ee
Thus according to the identification (\ref{X_x}), we find the
response function
  \be
  \chi(\omega, z)=\frac{1}{\omega
  r_b^4 T_3 S}\frac{H^{(1)}_{\frac2z+\frac12}\big(\frac{\omega}{zr_b^z}\big)}{H^{(1)}_{\frac2z-\frac12}\big(\frac{\omega}{zr_b^z}\big)}
  \, .
  \ee
It is then instructive to examine the
low-frequency behavior of the response function, expressed in the
form
 \be
  \chi (\omega, z)=\frac1{m(z)\, (i\omega)^2+\mu (\omega,z)}
  \, ,\label{chi_omega}
  \ee
 where $m$ is  an inertial mass  and  the $\mu$ term is the self-energy.
The low-frequency expansion, i.e. $\omega\ll r_b^z$, gives
  \be
  m(z) =\frac{T_3 S}{(4-z)r_b^{z-4}} \, ,\qquad\qquad\mu(\omega,z) =
  \gamma(z) (-i\omega)^{1+\frac4z}+ \delta (z) (-i \omega)^4+\cdots
  \ee
  with
  \be
  \gamma
  (z)=\frac{T_3 S}{(2z)^{4/z}}\frac{\Gamma(\frac{1}2-\frac2z)}{\Gamma(\frac12+\frac2z)}
  \, , \qquad\qquad\delta (z) =-\frac{T_3 S}{(4-3z)(4-z)^2r_b^{3z-4}}
  \, .
  \ee
 To avoid the breakdown of a low-frequency expansion  near $z=4/3$ and $z=4$, we have to restrict the value of $\omega$ such that in the expansion, the next order correction can not be larger than the order of interest. This restriction imposes a condition, $\omega < \left\vert (z-4) (z-4/3)\right\vert r_b^z$.
Apart from $z=4/3$ and $z=4$, the $\gamma$ term with a frequency
dependence $\omega^{1+\frac4z}$ will give the damping effect  on the
mirror. Additionally, the self-energy $\mu$
has a term proportional to $\omega^4$, which
is the next order result in a small $\omega$ expansion. As in the
case of the Brownian particle~\cite{Tong_12}, the similar
nonanalytic term in the self-energy, which has the
power-law dependence on frequency, is also found.
Since the object of
 interest is
 a two-dimensional mirror, as compared with a point particle in~\cite{Tong_12},
 the  power of the $r$
dependence in action $ S_{DBI} $  is increased by 2 to account for
the additional degrees of freedom. As a result, the critical value
 shifts to $z=4$.
%At the critical value $z=4$, the response function behaves like: \be
% \chi(\omega)\propto\frac1{(i\omega)^2+\frac{2i}{\pi}(i\omega)^2\ln\frac{e^{\gamma_E}\omega}{2zr_b^z}}
 % \ee}}
It has also been discussed in~\cite{Tong_12} that in spite that both
$m$ and
 $\gamma$  are changed from positive to negative values when $z$ goes from $1<z<4$ to $z>4$, its ratio $\gamma /m$ remains positive in a way that they still
give
 sensible results for describing the dynamics of the mirror.

Before closing this section, let us compare the $z=1$ case with
the field theoretic calculations in~\cite{Lee_05}, where the
mirror is coupled to a relativistic free scalar field. Here when
$z=1$, the self-energy term obtained from a holographic approach
is given by\be \mu (\omega,z=1)=T_3 S\bigg(\frac{\omega^4}{9 \,
r_b^{-1}}+\frac{i \omega^5}{6} +...\bigg)\, .\ee with an
ultraviolet energy cutoff $r_b$. The dominant terms in a small
$\omega$ expansion have the same $\omega$ dependence in both
cases, but different coefficients. The difference in the
coefficients may lie in the fact that the environment assumed in
the holographic approach is a strongly coupled field rather than a
free field in the field-theoretic approach. In particular, we
observe that the coefficient of the $\gamma$ term is proportional
to $T_{3}$, so it will increase in accordance with the coupling
constant $\lambda$ of the corresponding strongly coupled boundary
field. The connection that $T_3\propto\lambda$ merely reflects the
fact that the 3-brane tension $T_3$ is proportional to
$\alpha'^{-2}$ that in turn can be related to the coupling
strength $\lambda$ by $\lambda=L^4/\alpha'^2$ via AdS/CFT
correspondence where $L$ is the curvature radius in the Lifshitz
background.

\section{General dimension mirrors and the fluctuation-dissipation theorem}
We now generalize our previous results to an $n$-dimensional
mirror with its bulk counterpart as a $n+1$-brane with the
coordinates, $x^{I_n}(t,r,x_1,x_2..,x_n)$, where $I_n=n+1,...,d-1$
are the directions normal to the brane. Under the same assumptions
used in the 3-brane case, when the mirror moves along one of the
$I_n$ directions, the corresponding equation of motion  for
$\langle x \rangle$ is given by
  \be
  \label{mirror n}
 \frac{\partial}{\partial r}\bigg(r^{z+n+3}\frac{\partial}{\partial
 r} \langle  x \rangle  \bigg)+\frac{\omega^2}{r^{z-n-1}}\langle
 x\rangle=0  \,.
  \ee
Assuming the coupling described by a similar surface integral of
the stress tensor of the fields  as in
(\ref{coupling}), we then have the response function of an
$n$-dimensional mirror given by
 \be
 \label{response function}
  \chi_n (\omega, z)=\frac{1}{\omega r_b^{n+2} T_{n+1} S_n}\frac{H^{(1)}_{\frac{n+2}{2z}+\frac12}(\frac{\omega}{zr_b^z})}{H^{(1)}_{\frac{n+2}{2z}-\frac12}(\frac{\omega}{zr_b^z})}\,,
  \ee
 where $S_n$ is the mirror's surface area and $T_{n+1}$ is the $(n+1)$-brane tension.
 In the  low frequency limit,
the response function can be cast in the form
 \be \label{chin}
  \chi_n (\omega, z)=\frac1{m_n (z) (i\omega)^2+\mu_n (\omega,z)} \,
  ,
  \ee
  in which
  \be \label{m_mu_n}
  m_n (z)=\frac{T_{n+1} S_n}{(n+2-z)r_b^{z-n-2}},\qquad\qquad\mu_n(\omega,z)=\gamma_n(z)(-i\omega)^{1+\frac{n+2}z}+\delta_{n}(z) (-i\omega)^4+...
  \ee
with \be \label{gamman} \gamma_n (z)=\frac{T_{n+1}
S_n}{(2z)^{(n+2)/z}}\frac{\Gamma(\frac12-\frac{n+2}{2z})}{\Gamma(\frac12+\frac{n+2}{2z})}
\, , \qquad\quad\delta_n(\omega, z)=-\frac{T_{n+1}
S_n}{(n+2-3z)(n+2-z)^2r_b^{3z-n-2}} \, .\ee Here the critical value
is changed to $z=n+2$ as expected. The low-frequency expansion is
valid as long as $\omega <\left \vert [z-(n-2)]
[z-(n+2)/3]\right\vert r_b^z$.

In the following, we will examine the long time
dynamics of the mirror, in particular  its saturation mechanism on
velocity fluctuations.

\subsection{Fluctuation-dissipation theorem}
The stochastic behavior of the mirror  is reflected by the
two-point correlation function of its position. The idea of the
holographic duality is to relate the two-point function of
mirror's positions to the correlation function of
 the position of $n+1$-brane evaluated on the boundary of the bulk.
The fluctuations associated with the mirror's position in the
holographic setup result from the fluctuations around the brane vacuum state
in the bulk. In what follows, we quantize the modes normal to the
$n$-dimensional mirror surface. The procedure of the canonical
quantization mainly follows that in \cite{Holographic QBM}.

We first find the momentum conjugated to the coordinate $x$, which
describes the motion of the brane, from $S_{DBI}$, straightforwardly
generalized from (\ref{s_dbi}), as \be \pi
(t,r)=\frac{T_{n+1}}{r^{z-1-n}} \dot{x} (t,r) \, , \ee
for a rigid mirror so that $x$ does
not depend on $x_1,x_2,...,x_n$. The mode expansion on the position
operator ${x} (t,r) $ in its frequency space is given by \bea
\label{modes}
{\qquad}{x} (t,r) &=& \int_{-\infty} ^\infty \frac{d\omega}{\sqrt{ 2\pi}} \, x_{ \omega}(r) \, e^{-i \omega t}\, \nonumber\\
&=&\int_0^{\infty} d\omega \, U_{ \omega} (r) \bigg( a_{\omega} \,
e^{-i \omega t} + a^{\dagger}_{\omega} \, e^{i \omega t} \bigg) \, .
\eea The equal-time commutation relations give the Wronskian condition
of the mode functions
$U_{ \omega} $ in the Lifshitz geometry,
 \be
 \label{normalization}
 -i T_{n+1} S_n \int_0^{r_b} dr \frac{1}{r^{z-1-n}}\biggl\{U_{\omega}(r)e^{-i\omega t}\partial_t \left[ U_{\omega}(r) e^{i\omega t}\right]-\partial_t \left[ U_{\omega}(r) e^{-i\omega t}\right]U_{ \omega}(r)e^{i\omega t}\bigg\}=1\, .
 \ee
The vacuum state is annihilated by $a_{\omega}$ for
all $\omega$ modes. The mode functions are solved
from (\ref{mirror n}) with the Neumann boundary condition
$x'_n(r_b,t)=0$ and the Wronskian
condition~(\ref{normalization}). Thus the two-point function associated
with  the rigid $D$-brane fluctuations at $r=r_b$ is obtained as
 \bea
 \label{correlator}
  \langle X_{\omega}\,  X_{-\omega} \rangle &=& \langle x_{\omega}(r_b)\,  x_{-\omega}(r_b) \rangle = 2\pi U^2_{\omega} (r_b) \nonumber\\
  &=& \frac{4 z r_b^{z-2-n}}{\pi \omega^2 T_{n+1}
  S_n}\left[J^2_{\frac{n+2}{2z}-\frac12}\bigg(\frac{\omega}{zr_b^{z}}\bigg)+Y^2_{\frac{n+2}{2z}-\frac12}\bigg(\frac{\omega}{zr_b^z}\bigg)\right]^{-1}\,.
   \eea
 The fluctuation-dissipation theorem relates the fluctuations of the mirror's position to the imaginary part of its response function in (\ref{response function}), and can be shown to take the form
   \be
\langle X_{\omega} X_{-\omega}\rangle=2 \mbox{Im}\chi_n(\omega , z)
\, .
   \ee
Evidently the bulk results can really capture the essential
properties of Brownian motion in a general
environment.

\subsection{Supraohmic behavior and velocity fluctuations}
According to the general  Langevin equation (\ref{langevin_eq}), the
effects of the environment on the system are classified according to the
low-frequency behavior of the imaginary part of the self-energy
$\mu$, i.e. the friction term of the general form $\gamma
\omega^{k+1}$, as subohmic, ohmic, and supraohmic
for $k<0$, $k=0$ and $k>0$ respectively. The low-frequency expansion
of the response function in (\ref{response
function}) from the holographic approach gives $k=({n+2})/{z}$, and for the
positive value of $n$ and $z$ it corresponds to the
supraohmic case. As a result of the
fluctuation-dissipation relation, it is expected that the properties of the fluctuations associated with the mirror will be
different from the ohmic environment, such that they result in the
different mechanism for the evolution of velocity fluctuations
toward their saturation.

The long-time dynamics of velocity fluctuations relies on the
behavior of the correlation function in the low-frequency limit,
obtained from (\ref{correlator}),  as $\langle X_{\omega}\,
X_{-\omega} \rangle\sim\omega^{\frac{n+2}{z}-3}$
for $1<z<n+2$ and $\langle X_{\omega}\,
X_{-\omega}\rangle\sim\omega^{-\frac{n+2}{z}-1}$
for $z>n+2$. Thus from the
Langevin equation (\ref{langevin_eq}), the correlation function of
the random forces is found to be
 \be
\langle R_{n;\omega}R_{n;-\omega}\rangle=\frac{\langle X_{\omega}
X_{-\omega}\rangle}{\chi_n(\omega,z)\chi_n(-\omega,z)} \, .
 \ee
 Then, together with the results in~(\ref{chin}), (\ref{m_mu_n}), and
 (\ref{gamman}), we have
  \be\langle
R_{n;\omega}R_{n;-\omega}\rangle\sim
\omega^{\frac{n+2}{z}+1}\, , \ee
 in the low-frequency limit.
Thus, \be \int_{-\infty}^{\infty}  d\tau \langle R_n
(\tau)R_n(\tau')\rangle =2 \int_{0}^{\infty}  d\tau \langle R_n
(\tau)R_n(\tau')\rangle = \int_{-\infty}^{\infty}  d\tau \, \eta
(\tau-\tau') \propto\lim_{\omega\rightarrow 0}\langle R_{n;\omega}
R_{n;-\omega} \rangle = 0\,. \ee We show that the integration of the
force-force correlation function over the whole time regime is found
vanishing. In general, the positive force-force correlation may
contribute to the growth of the velocity dispersion, whereas the
negative correlation may halt its growth. The cancelation between
them implies that the velocity dispersion will reach a constant at
asymptotical times. It certainly leads to a rather different
saturation mechanism than  an ohmic environment where the
force-force correlation function remains positive at all times. This
scenario for the supraohmic case
 has been discussed in~\cite{Lee_05,Lee_13,Lee_09}. Here we extend
our precious study by considering the strongly coupled
environment in quantum critical theories.

Velocity fluctuations can be computed straightforwardly as follows:
  \be
  \label{velocity}
  \langle(\delta v(t))^2\rangle=\langle v^{2}(t)\rangle-\langle v(t)\rangle^2
  =\int\frac{d\omega}{\pi}\,\omega^2 \langle X_{ \omega} X_{-\omega}\rangle\, (1-\cos\omega t)
  \,.
  \ee
The saturated {value of the velocity dispersion} is
  \be
  v_s^2 = \int\frac{d\omega}{\pi}\,\omega^2 \langle X_{ \omega} X_{-\omega} \rangle
  = 4 z^2  {\cal{N}} \frac{r_b^{2z-2-n}}{T_{n+1} S_n}  \,,
  \ee
  with
  \be
  {\cal{N}}=\int_0^{1/2z} d y \left(J^2_{\frac{n+2}{2z}-\frac12}\big(y\big)+Y^2_{\frac{n+2}{2z}-\frac12}\big(y\big)\right)^{-1}\, ,\ee
where we impose the frequency cutoff $r_b^z$. Using the expression of the inertial mass
in~(\ref{m_mu_n}), we have $m_n v_s^2 \sim r_b^z$ as expected on
dimensional grounds. Nevertheless, the late-time saturation behavior of the velocity fluctuations
 follows the power law.
We find that for $z>n+2$,
 \be
\langle(\delta v(t))^2\rangle-  v_s^2 \, \propto \, -\frac{1}{
T_{n+1} S_n}(2z)^{\frac{n+2}{z}}t^{\frac{n+2}z-2} \, ,
 \ee
and for $1< z< n+2$,
  \be
\langle(\delta v(t))^2\rangle-  v_s^2 \,\propto \, -\frac{1}{T_{n+1}
S_n}
\frac{(2z)^{-\frac{n+2}{z}+2}}{r_b^{2(2+n-z)}}t^{-\frac{n+2}{z}} \,
.
  \ee
They are our main results in this paper. Notice that
different power-law behavior in time is mainly due to the fact that the low frequency behavior of
the response function is dominated, respectively, by the inertial mass
term for $ 1< z < n+2$ and the $\gamma$ term for $z> n+2$. The
$r_b$ dependence can also be realized from the dimensional argument,
and can be substituted by the inertial mass through the
relation~(\ref{m_mu_n}), which carries units with dimension
$[m]=2-z$ in this problem.

\section{Hawking radiation and Thermal motion}
We now heat up the environment in this holographic model with a
Lifshitz black hole background. We will study the response
function and thermal fluctuations for a $n+1$-brane in this
background, and explicitly verify the corresponding
fluctuation-dissipation theorem.

The background metric of a Lifshitz black hole in $d+1$ dimensions
is
  \be
  \label{lifshitz bh}
  ds^2=-r^{2z}f(r)dt^2+\frac{dr^2}{f(r)r^2}+r^2d\vec{x}^2 \, .
  \ee
In the low-frequency limit, the actual form of the function $f(r)$
is irrelevant, but the function is required to satisfy the
properties like $f(r)\rightarrow1$ for $r\rightarrow\infty$ and
$f(r)\simeq c(r-r_h)$ near the black hole horizon $r_h$ with
$c=({d+z-1})/{r_h}$. The temperature of the black hole and also of
the boundary field theory is \be \label{r_h}
\frac1T=\frac{4\pi}{d+z-1}\frac1{r_h^z}\, . \ee The equation of
motion for the expectation value of $x_{ T}$ moving along one of
the directions normal to the $(n+1)$-brane in this background
becomes
 \be
  \label{mirror n with T}
 \frac{\partial}{\partial r}\biggl( r^{z+n+3}f(r)\frac{\partial }{\partial
 r} \langle x_{T} \rangle \biggr)+\frac{\omega^2}{r^{z-n-1}f(r)}\langle x_{ T} \rangle=0 \,
 .
  \ee
This equation of motion can be cast into the Schrodinger-like
equation using tortoise coordinate $r^*=\int drf(r)^{-1}r^{-z-1}$ as
follows:
 \be
\frac{d^2y_T}{dr^{*2}}+\Bigl[\omega^2-V(r)\Bigr]y_T=0 \, ,
 \ee
where $y_T=r^{\frac{n}{2}+1}x_T$ and
$V(r)=r^{2z}(\frac{n}2+1)f(r)\left[(z+1+\frac{n}2)f(r)+rf'(r)\right]$.
We first find the solutions in three separate regimes
specified below. The whole solution will be obtained by the
matching method (see \cite{Tong_12} and references therein). Here we
just summarize the final results.

In regime (A), the near-horizon region, defined by $r\rightarrow r_h$ and thus
$V(r)\ll\omega^2$,
 the solution in the small $\omega$ approximation is given by
 \be
 \label{regionA}
x^{(A)}_{T}\simeq D_T
\left[1-\frac{i\,\omega}{cr_h^{z+1}}\ln(r-r_h)+O(\omega^2)\right] \, ,
 \ee
where $D_T$ is a constant and the in-falling boundary condition at
$r=r_h$ is chosen. In regime (B), where $r$ takes intermediate
values and $V(r)\gg\omega^2$, the  solution is found as
  \be
  x^{(B)}_{T}\simeq D_T (1-i\,\omega r_h^{n+2}k)\biggl[1+\mathcal{O}(\omega^2)\biggr]+i\,D_T \omega
  r_h^{n+2}\int_r^{\infty}\frac{dr'}{r'^{z+n+3}f(r')}\biggl[1+\mathcal{O}(\omega^2)\biggr] \, ,
  \ee
where $\kappa$  is an $\omega$-independent integration constant.
Finally in regime (C), as $r\rightarrow r_b$, the equation reduces
to the one in Lifshitz geometry, and its solution
is the Bessel functions. In the small $\omega$ limit, we have the
expansion
   \bea
   \label{regionC}
   x^{(C)}_{T}&\simeq & i\,\frac{D_T}{z+n+2} \frac{\omega
  r_h^{n+2}}{
  r^{n+2+z}}\biggl[1-\frac1{\frac{n+2}{2z}+\frac32}\big(\omega/2zr^z\big)^2+\mathcal{O}(\omega^4)\biggr]
\nonumber\\
&& \quad \quad + D_T (1-i\,\omega
  r_h^{n+2}\kappa )
  \biggl[1+\frac1{\frac{n+2}{2z}-\frac12}\big(\omega/2zr^z\big)^2+\mathcal{O}(\omega^4)\biggr]
   \, . \eea
Thus the response function is given by
  \be
  \chi_{ n T}(\omega)=\frac{x^{(C)}_{T} (r_b,\omega)}{T_{n+1} S_n
  r_b^{z+n+3}x'^{(C)}_{T}(r_b,\omega)}=
  \frac1{m_{n T}(z) (i\omega)^2-\gamma_{n T}(z) i\omega}+O(\omega)\,,
  \ee
where \be \label{m_gamma_T} m_{n
T}(z)=\frac{T_{n+1}S_n}{r_b^{z-n-2}}\biggl\{
\frac1{n+2-z}+\bigg(\frac{r_h}{r_b}\bigg)^{2n+4}\Bigl[(n+2+z)-\kappa
r_b^{z+n+2}\Bigr]\biggr\}\, , \, \gamma_{n T}(z)=T_{n+1}S_n r_h^{n+2}
\,. \ee  The inertial mass $m_{ n T}$ and the
damping coefficient $\gamma_{nT}$ have the temperature dependence
through the black hole temperature~(\ref{r_h}). Since the damping
term has linear $\omega$ dependence, the stochastic dynamics of
the mirror in the thermal environment will be
expected to be ohmic.

Next we
quantize the modes in this thermal background. We use the approximate solutions found above and impose the Neumann
boundary condition at $r=r_b$. Since the mode function near the
horizon $r=r_h$ exhibits logarithmic divergence, an infrared energy
cutoff scale $\epsilon$ as $ r \rightarrow r_h$ is introduced for
regularization. {Without proper renormalization,  the result for the
counterpart of~(\ref{correlator}), denoted as $\langle
 X_{T;\omega}X_{T;-\omega}\rangle$, can be pathological in} the background of
Lifshitz black hole. We may absorb this infrared divergence by
carefully defining the density of states  as $\Delta\omega=4\pi^2
T/\ln(1/\varepsilon)$~\cite{Holographic QBM}. The modes expansion now
becomes \bea \label{modes_T}
{x}_{ T} (t,r) &=& \int_{-\infty} ^\infty \frac{d\omega}{\sqrt{ 2\pi}} \, x_{ T\, \omega}(r) \, e^{-i \omega t}\, \nonumber\\
&=&\sqrt{\frac{\ln(1/\varepsilon)}{4\pi^2 T}} \, \int_0^{\infty}
d\omega \, U_{T \omega} (r) \bigg( a_{\omega} \, e^{-i \omega t} +
a^{\dagger}_{\omega} \, e^{i \omega t} \bigg) \, . \eea The
corresponding Wronskian condition is
 \be \label{norm_T}
 -i T_{n+1} S_n \int_{r_h+\varepsilon}^{r_b} dr \frac{1}{f(r)r^{z-1-n}}\biggl\{U_{T \omega}(r)e^{-i\omega t}\partial_t \Bigl[ U_{T \omega}(r) e^{i\omega t}\Bigr]-\partial_t \Bigl[ U_{T \omega}(r) e^{-i\omega t}\Bigr]U_{T \omega}(r)e^{i\omega t}\biggr\}=1\,
.
 \ee
It is quite straightforward to find the complete solution over three
separate regimes using the matching method. Imposing the Neumann
boundary condition at $r=r_b$ gives a solution with an undetermined
constant $D_T$ that can be fixed by the
Wronskian condition
in~(\ref{norm_T}). Here we only keep the divergent
parts of this integral as $r_b\rightarrow\infty$ and
$\varepsilon\rightarrow 0$. Thus, in the small $\omega$
approximation, the most relevant terms in the integral of
(\ref{norm_T}) come from solutions in regions (A) and (C), which take the forms,
 \be
 U^{(A)}_{T, \omega} (r)\simeq D_T \biggl[\frac{(n+2)^2-z^2}{(2z)^2}(\omega/2zr_b^z)^{\frac{n+2}{z}-1}+\frac{z+n+2}{d+z-1}(\omega/2z
 r_h^z)^{\frac{n+2}{z}+1}\ln(r-r_h)\biggr] \, ,
 \ee
 and
   \be
   U^{(C)}_{T, \omega} (r) \simeq D_T \biggl[\frac{(n+2)^2-z^2}{(2z)^2}(\omega/2zr_b^z)^{\frac{n+2}{z}-1}
   \Bigr)\biggr] \, .
   \ee
The undetermined constant $D_T$ is then given by
\be |D_T|^2=\frac1{2\omega T_{n+1} S_n}
 \biggl[\frac{(n+2)^2-z^2}{(2z)^2}(\omega/2zr_b^z)^{\frac{n+2}{z}-1}
 \biggr]^{-2}
 \biggl[\frac{r_h^{n-z+2}}{d+z-1}\ln(1/{\varepsilon})\biggr]^{-1} \, ,
  \ee
which apparently suffers from
the $\ln(1/{\varepsilon})$ divergence. Note that in the  Lifshitz
black hole background, the modes of the $D$-brane in the bulk get
excited to obey the thermal distribution $\langle
a_{\omega}a^{\dag}_{\omega}\rangle=(1-e^{-\frac{\omega}{T}})^{-1}$.
Putting all together, the leading term in the small $\omega$
expansion of $\langle X_{ T,\omega} X_{T,-\omega} \rangle$ is
obtained from the mode function $U^{(C)}_{T, \omega} (r)$ evaluated
at $r=r_b$ as  \bea
   \langle X_{
T,\omega} X_{T,-\omega} \rangle   = \langle x_{ T;\omega}(r_b)\,
x_{T;-\omega}(r_b) \rangle &=& \frac{2\pi}{1-e^{-\frac{\omega}{T}}}
\frac{\ln(1/\varepsilon)}{4\pi^2 T} \vert U^{(C)}_{T \, \omega}
(r_b) \vert^2  \nonumber\\
  &\simeq&
  \frac1{1-e^{-\frac{\omega}{T}}}\frac2{T_{n+1}S_n\omega}r_h^{-n-2}
  \, ,
   \eea
which is divergence free.

The fluctuation-dissipation theorem in the thermal
environment,
  \be
 \langle
 X_{T;\omega}X_{T;-\omega}\rangle=2\left(\frac1{1-e^{-\frac{\omega}{T}}}\right)\mbox{Im}\chi_{n T} (\omega)
 \, ,
  \ee
can be checked explicitly from the above results in their
low-frequency limit. Additionally, we find that
 \be \langle
R_{n T;\omega}R_{n T;-\omega}\rangle=\frac{\langle X_{T;\omega} X_{
T;-\omega}\rangle}{\chi_{n T} (\omega,z)\chi_{n T} (-\omega,z)} \,
\approx \frac{\langle X_{ T;\omega} X_{T;-\omega}\rangle}{\omega^2
\gamma^2_{n T}} \approx T_{n+1} S_n \, 2 T \, \left(\frac{4\pi
T}{d+z-1}\right)^{n+2}
 \ee
using the above expressions of  $\gamma_{n T}$ in~(\ref{m_gamma_T})
and $r_h$ in~(\ref{r_h}). Thus, this in turn can be translated into
the white noise forces in the time domain, \be \langle R_{n T}
(t) R_{n T} (t')\rangle = T_{n+1} S_n \, 2 T \, \left(\frac{4\pi
T}{d+z-1}\right)^{n+2} \delta (t-t') \, . \ee As a result, velocity
fluctuations is anticipated to evolve in the same way as in the
Brownian motion in the ohmic case. They increase initially as the consequence of the energy input from the noise forces. When $ t \sim \gamma_{n
T}^{-1}$, the damping effect comes into the play, and then slows down
their growth into saturation. The saturated value of velocity
fluctuations is
 $
  v_{T s}^2 \approx T/m_{n T}^2 \, ,$
  where $m_{n T}$ is defined in~(\ref{m_gamma_T}).
The relaxation dynamics follows an exponential behavior in time
within a time scale determined by $\gamma_{nT}$
in~(\ref{m_gamma_T}). The main finding of this paper is to obtain
the general result of damping in a strong coupling environment by
\be \gamma_{n T}(z)=T_{n+1}S_n
\left(\frac{4\pi T}{d+z-1}\right)^{n+2}\, , \ee where
{from~(\ref{r_h}) we replace $r_h$ by $T$ to explicitly show the
temperature dependence of the result.

In particular, for $z=1$ (the relativistic environmental field) and
for a two-dimensional mirror, \be \gamma_T =T_3 S \left(2\pi
T/d\right)^{4} \, .\ee The ohmic dynamics and the $T^4$ dependence
of $\gamma_T$ are in agreement with the findings in~\cite{Lee_05},
but the {proportionality} constant is different {between the
strong coupling environment and} the free field background as
expected. Notice that in the strong field theory with coupling
 $\lambda>>1$, it will always show enhancement, since
$T_{n+1}\propto\lambda^{\frac{2+n}4}$ following the similar
arguments in the 3-brane case.

\section{Summary and outlook}
In this paper, we have successfully established the holographic
setup for the nonequilibrium dynamics of a moving mirror coupled
to quantum critical theories. The aim of this work is to
understand the quantum microphysics of nonequilibrium statistical
problems via holographic duality. The mirror with perfect
reflectance is realized by a $n+1$-brane of the bulk theory in the
Lifshitz geometry. The excitations of the bulk brane, due to
either its vacuum state in the Lifshitz geometry or the thermal
state in the Lifshitz black hole background, render the mirror
undergoing Brownian motion. Nevertheless, they exhibit rather
different damping behaviors, in particular, on the evolution of
velocity fluctuations. The dissipation exerted on the mirror in
the vacuum case is found to be supraohmic. For an initial growth
of velocity fluctuations, the saturation at late times follows the
power-law: when $ z> n+2,$ the saturation behavior is like $
t^{\frac{n+2}z-2},$ and when $1< z< n+2,$ $ t^{-\frac{n+2}{z}}$,
respectively. On the contrary, in the Lifshitz black hole
background, the dissipation caused by thermal excitations
 becomes ohmic  so that the relaxation dynamics toward
saturation is exponentially fast with a relaxation time scale
$ \propto 1/\gamma_{nT}$ where $\gamma_{n
T}(z)=T_{n+1}S_n \left[(4\pi T)/(d+z-1)\right]^{n+2}.$ In the small
displacement approximation, for the relativistic quantum field
($z=1$) and a two-dimensional mirror, the dissipation/relaxation
behavior of the mirror influenced from the quantum field
via the holographic approach follows the same dynamics as is
obtained by the field theoretic approach. However, all results based
upon the holographic duality are enhanced by the brane tension $T_n$
to account for the strong coupling effects of the environment.

Finally,  we would like to  point out some of our future work. In
view of a close relation between the holographic approach and the
field-theoretical study of the nonequilibrium problems
in the linear response regime, the generalized Langevin
equation in~(\ref{langevin_eq}), which can be derived from the known
interactions between the system and the bath via the method of
influence functional, may be obtained from holography~\cite{Son_03}. It
is then an important next step to establish this correspondence more
explicitly {by} studying the full nonequilibrium dynamics {in} a
strong coupling environment beyond the linear response.

\begin{acknowledgments}
This work was supported in part by the National
Science Council, Taiwan.
\end{acknowledgments}

\end{document}